\documentclass[pra,twocolumn,tightenlines,showpacs,nofootinbib]{revtex4}
\usepackage{bm,dcolumn,amsmath,graphicx,amsfonts,amssymb}
\usepackage{epsfig}

\begin{document}
\title{Comment on ``Reappraisal of the Electric Dipole Moment
  Enhancement Factor for Thallium''} 

\author{V. A. Dzuba and V. V. Flambaum}
\affiliation{School of Physics, University of New South Wales,
Sydney, NSW 2052, Australia}

\date{ \today }

\begin{abstract}
Recent paper by Nataraj {\em et al} (Phys. Rev. Lett. {\bf 106},
200403 (2011)) presents calculations of the EDM enhancement factor for Tl,
which disagrees with previous most accurate calculations.
The authors claim that their calculations of Tl EDM are the most accurate 
due to more complete treatment of higher-order correlations. In this
note we argue that this claim is not supported by sufficient
evidence. Nataraj {\em et al} also present misleading comments about
our calculations. We explain our method and reply to the Nataraj {\em
  et al} comments.  
\end{abstract}
\maketitle

Recent paper~\cite{Nataraj} presents
calculations of the EDM enhancement factor for Tl, which disagrees
with previous most accurate 
calculations~\cite{Liu,Dzuba09}.  
The calculations of this
kind are used for a search for new physics beyond the standard model
in atomic  experiments. 
Therefore, it is important to use the most accurate results for the
interpretation of the experiments.


The authors of~\cite{Nataraj} make many misleading comments about
our calculations of thallium EDM~\cite{Dzuba09}. 
Contrary to the statement in~\cite{Nataraj} the atomic
electric field interacting with electron EDM is 
calculated in \cite{Dzuba09} as a derivative of the total
potential which includes both nuclear and electron parts. 
This is done in a same way as in our early work on EDM of Fr and
Au~\cite{BDFM}. The formula $\mathbf{E} = Ze\mathbf{r}/r^3$ for the
leading contribution to the atomic electric field presented on first
page of~\cite{BDFM} may indeed make an impression 
that only nuclear field is included. However, the formula (2) few
lines below clearly includes screening functions $Q(r)$ and $P(r)$ for
both the nuclear Coulomb and the external electric field. 
By the way, the inclusion of the electron electric field change the
matrix elements of the electron EDM for thallium by 0.4\% only. This
is because the main contribution comes from short distances where the electron
electric field is small since the electron potential rapidly tends to
a constant inside the $1s$ orbital.


\begin{figure}
\epsfig{figure=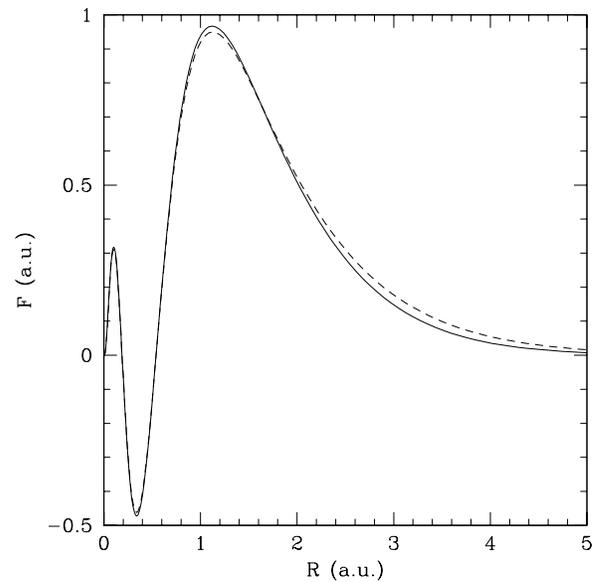,scale=0.40}
\caption{The $5d_{5/2}$ core function of Tl calculated in $V^{N-3}$
  (solid line) and $V^{N-1}$ (dashed line) approximations.}
\label{fig:ff}
\end{figure}

The authors of~\cite{Nataraj} claim that atomic core is strongly
contracted in the $V^{N-3}$ zero approximation used in our
calculations~\cite{Dzuba09}.  In fact, it is not. 
Fig.\ref{fig:ff} shows the outermost $5d_{5/2}$ core function 
of Tl calculated in $V^{N-3}$ and $V^{N-1}$ approximations. The
difference between the functions is very small. 
This is due to the fact that the valence $6s$ and $6p$ electrons 
are located outside of atomic core. Their charge
distribution creates almost constant potential and no electric field
inside the sphere where all inner electrons are located. Therefore,
the valence electrons have practically no effect on the core wave functions
(see~\cite{VNM} for a detailed discussion).  
The change is even smaller for other core functions. 
The core functions enter the
configuration interaction (CI) Hamiltonian  via core potential
$V_{core}$ to which all core electrons contribute (from $1s$ to
$5d$). The difference for 
$V_{core}$ in $V^{N-3}$ and $V^{N-1}$ approximations is very small~\cite{VNM}.
Moreover, the corresponding corrections to the configuration
interaction (CI) Hamiltonian have been included in~\cite{Dzuba09}
using the many-body perturbation theory approach.  

As it is well known, the eigenstates of a Hamiltonian do not depend on
the basis one uses.
The valence states are indeed different in the $V^{N-3}$ and $V^{N-1}$
approximations. However, this should have no effect on the final
results as long as the saturation of the basis for valence states is
achieved. There are 
only two conditions the basis states must satisfy: (a) they must be
orthogonal to the core, and (b) they must constitute a complete set
of states. Therefore, it does not matter whether valence states are
calculated in the $V^{N-3}$ or $V^{N-1}$ potential or by any other method
(e.g., a popular basis $\psi_n(r)=r^n\psi_0(r)$~\cite{KPF96}), the final
results should be the same. If there is any difference in the results, the
most likely reason for this is the incompleteness of the basis set. 

In spite of no difference in final results there is a good reason
for the use of the $V^{N-3}$ approximation -- the simplicity and good
convergence of the many-body perturbation theory (MBPT) for the
core-valence correlations.  When an approximation different from the
$V^{N-3}$  is used one has to include the so
called {\em subtraction diagrams}~\cite{DFK96}, while there are no such
diagrams in the $V^{N-3}$ approximation. Large energy denominators 
suppress the value of the correlation terms in the $V^{N-3}$
approximation ensuring good convergence of the MBPT~\cite{VNM}.
There must be large cancelation between subtraction and other
diagrams to ensure the same final results if any other initial
approximation is used. This is very similar to the well known fact that
the Hartree-Fock basis is the best choice for any MBPT
calculations. Initial approximation might be better in some other
approximation, however, strong cancellations between the subtraction and
other diagrams would lead to poor convergence of the MBPT.

The authors of~\cite{Nataraj} claim that the major drawback of our
work~\cite{Dzuba09} is the inclusion of the core-valence correlations
in the second order only. However,  the correlations between the valence
 electrons and core electrons below the $6s$ state are small which is evident
from the fact that their inclusion change the EDM of Tl by 3\%
only~\cite{Dzuba09}.
Therefore, only the 
correlations between three valence electrons should be treated
to all orders. This is done in~\cite{Dzuba09} to a very high precision
using the CI technique.

Early calculations of thallium EDM by Liu and Kelly~\cite{Liu} were
performed by the same 
relativistic coupled-cluster (RCC) method as those used
in~\cite{Nataraj}. Table~I of~\cite{Nataraj} presents term by term
comparison between the contributions to the enhancement factor calculated
in both works. For some terms the agreement is perfect, for others
there is strong disagreement. The authors of~\cite{Nataraj} claim that
the disagreement is due to more accurate treatment of higher-order
correlations in their work. It might be possible to prove this claim
by switching off the higher-order terms and reproducing the results
of~\cite{Liu}. 
The benefit of having the test is enormous. Without it, no other reasons
for disagreement can be excluded (e.g., incompleteness of the basis).
There is no indication that such test has been performed
in~\cite{Nataraj}. 

In conclusion, we would like to note that the calculations of the Tl
EDM  due to electron 
EDM~\cite{Nataraj} and SPS interaction~\cite{Sahoo08} do not satisfy a
simple consistency 
test: the ratio of the EDMs due to two operators must be approximately
equal to the ratio of the $s-p$ single-electron matrix elements of these
 operators. 
This ratio is the same for all important single-electron matrix
elements.
This is because only short distances, where
single-electron energies can be neglected, contribute to the
single-electron matrix elements of the T,P-odd operators.
The ratio approximately equals to $89 d_e/C^{SP} 10^{-18} \ e \
$cm according to analytical estimates, or $83 d_e/C^{SP}
10^{-18} \ e \ $cm in more accurate numerical
calculations~\cite{Dzuba11}.
This ratio must hold in any order of the MBPT if the
$s_{1/2}- p_{1/2}$ T,P-odd matrix elements dominate at the
Hartree-Fock level (see \cite{Dzuba11} for details).
The ratio of the results of \cite{Nataraj} and
\cite{Sahoo08} is $115 d_e/C^{SP} 10^{-18} \ e \ $cm. 
This may indicate that important many-body effects are missed in one
(or both) of the works~\cite{Nataraj,Sahoo08}.


\begin{thebibliography}{99}
\bibitem{Nataraj} H. S. Nataraj, B. K. Sahoo, B. P. Das, and D. Mukherjee,
  Phys. Rev. Lett. {\bf 106}, 200403 (2011). 
\bibitem{Liu} Z. W. Liu and H. P. Kelly, Phys. Rev. A {\bf 45}, R4210 (1992). 
\bibitem{Dzuba09} V. A. Dzuba and V. V. Flambaum,
      Phys. Rev. A {\bf 80}, 062509 (2009).

\bibitem{BDFM} T. M. R. Byrnes, V. A. Dzuba, V. V. Flambaum, and D. W. Murray,
Phys. Rev. A, {\bf 59}, 3082 (1999).

\bibitem{VNM} V. A. Dzuba, 
      Phys. Rev. A, {\bf 71}, 032512 (2005); V. A. Dzuba and V. V. Flambaum,
      Phys. Rev. A. {\bf 75}, 052504 (2007).

\bibitem{KPF96} M. G. Kozlov, S. G. Porsev, and V. V. Flambaum,
J. Phys. B {\bf 29}, 689 (1996).

\bibitem{DFK96} V. A. Dzuba, V. V. Flambaum, and M. G. Kozlov,
Phys. Rev. A, {\bf 54}, 3948 (1996).

\bibitem{Sahoo08} B. K. Sahoo, B. P. Das, R.K. Chaudhuri, D. Mukherjee,
and E. P. Venugopal, Phys. Rev. A {\bf 78}, 010501(R) (2008);{\bf 78},
039901(E) (2008).  



\bibitem{Dzuba11} V. A. Dzuba, V. V. Flambaum, and C. Harabati, to be
  published. 
\end{thebibliography}
\end{document}